\documentclass[reprint,superscriptaddress,amsmath,amssymb,aps,pra]{revtex4-2}

\usepackage{graphicx}
\usepackage{dcolumn}
\usepackage{bm}
\usepackage{dsfont}
\usepackage{physics}
\usepackage{soul}
\usepackage{braket}
\begin{document} 

\title{Mechanistic principles of exciton–polariton relaxation}

\author{Ian Haines}
\altaffiliation{equal contribution}
\affiliation{Department of Chemistry, Texas A\&M University, College Station, Texas 77843, USA}

\author{Arshath Manjalingal}%
\altaffiliation{equal contribution}
\affiliation{Department of Chemistry, Texas A\&M University, College Station, Texas 77843, USA}

\author{\\Logan Blackham}
\affiliation{Department of Chemistry, Texas A\&M University, College Station, Texas 77843, USA}

\author{Saeed Rahamanian Koshkaki}
\email{rahmanian@tamu.edu}
\affiliation{Department of Chemistry, Texas A\&M University, College Station, Texas 77843, USA}

\author{Arkajit Mandal}%
\email{mandal@tamu.edu}
\affiliation{Department of Chemistry, Texas A\&M University, College Station, Texas 77843, USA}

\begin{abstract}
Exciton-polaritons are light-matter hybrid quasi-particles that have emerged as a flexible platform for developing quantum technologies and engineering material properties. However, the fundamental mechanistic principles that govern their dynamics and relaxation remain elusive. In this work, we provide the microscopic mechanistic understanding of the exciton-polariton relaxation process that follows from an excitation in the upper polariton. Using both mixed quantum-classical simulations and analytical analysis, we reveal that phonon-induced upper-to-lower polariton relaxation proceeds via two steps: the first step is a vertical inter-band transition from the upper to the lower polariton, which is followed by a second step that is a phonon-induced Fr\"{o}hlich  scattering within the lower polariton. We find that in materials of finite thickness (which include filled cavities), phonon-induced polaritonic intraband Fr\"{o}hlich scattering is significantly suppressed. We show that the microscopic origin of this suppression is phonon-fluctuations synchronization (or self-averaging) due to the polaritonic spatial delocalization in the quantization direction. Finally, we show that the same phonon fluctuation-synchronization effect plays a central role across polaritonic relaxation pathways, and we derive simple analytical expressions that relate a material’s finite thickness to the corresponding relaxation rate constants.
\end{abstract}

\maketitle
\section{Introduction}
Exciton-polaritons, formed by the strong coupling of materials or organic molecules to quantized radiation inside an optical cavity~\cite{hutchison2012modifying, basov2020polariton, MandalCR2023, li2022molecular}, are an emerging platform that shows growing promise for application in classical computing~\cite{OpalaOME2023, SanvittoNM2016, ZasedatelevNP2019, AmoNP2010}, neuromorphic computing~\cite{BallariniNL2020, GhoshAQT2021, MirekNL2021}, quantum computing~\cite{RojasPRB2023, GhoshNPJQ2020, BerloffNM2017}, chemical catalysis~\cite{xiang2024molecular, nagarajan2021chemistry, thomas2019tilting, ahn2023modification, bhuyan2023rise,rashidi2025efficient}, and information transduction~\cite{manjalingal2025tilted, yang2023enabling}.
Despite extensive theoretical efforts to understand exciton–polaritons~\cite{tichauer2021multi, xu2023ultrafast, chng2025quantum, ying2025microscopic, krupp2025quantum, sokolovskii2025disentangling, blackham2025microscopic, ghosh2025mean, perez2025radiative, wang2025robust, catuto2025interplay, AroeiraJCP2025}, the microscopic mechanisms governing their dynamics remain poorly understood. A key limitation of many theoretical studies is the reliance on simplifying approximations that contradict experimental reality. Such approximations include the long-wavelength approximation, single-emitter approximation, single-layer approximation, and single cavity-mode approximation. Although these approximations may reproduce certain spectroscopic features, they often fail to capture key dynamical aspects of exciton–polaritons, leading to ambiguities or inaccurate descriptions of exciton–polariton dynamics.
\begin{figure}
\centering
\includegraphics[width=1.0\linewidth]{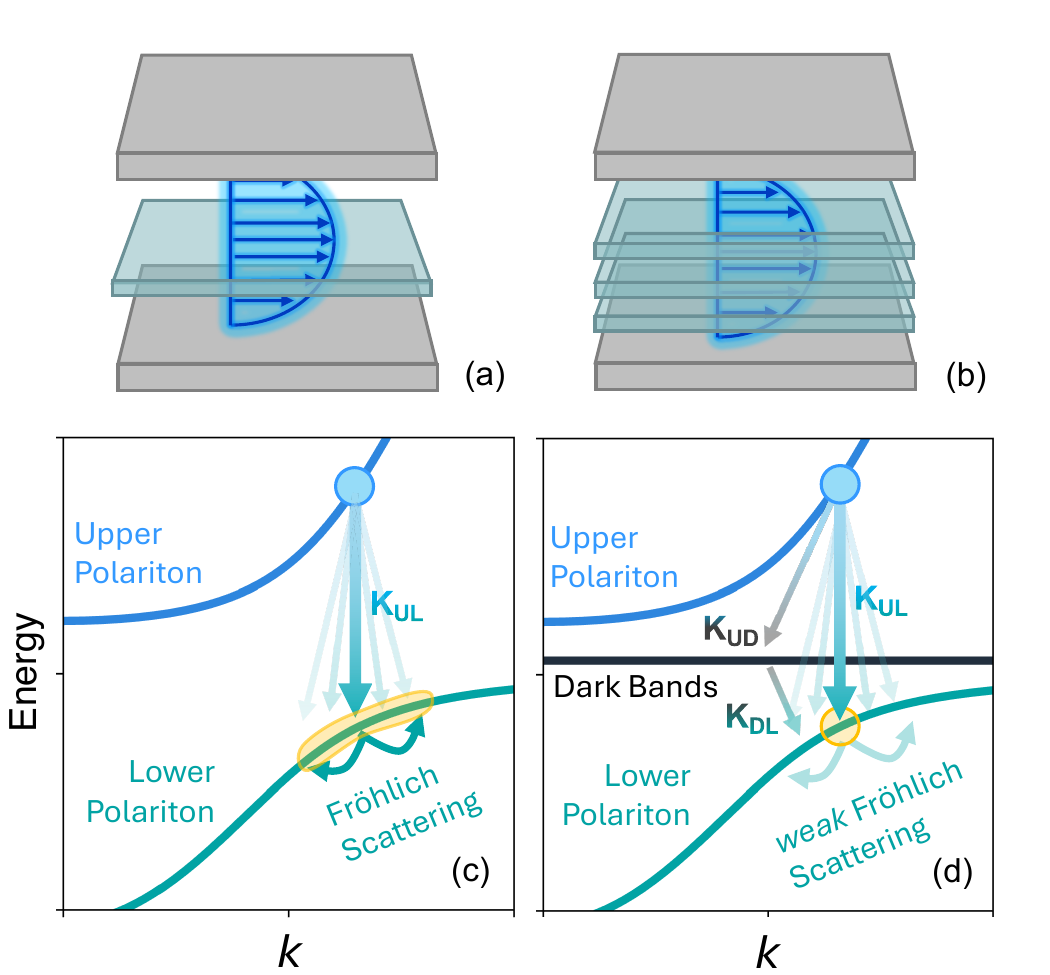}
\caption{\footnotesize Schematic of polariton dynamics. (a and b) Schematic drawing of a single-layer and multilayered material in an optical cavity. (c and d) Schematic drawing of the single-layer and multilayer band structures with energy on the $y$-axis and wavevector ($k$) along the horizontal axis. $K_{UL}$ ($K_{DL}$, $K_{UD}$) is the relaxation rate from the upper polariton (dark states, upper polariton) to lower polariton (lower polariton, dark states). }\label{fig1}
\end{figure}


Existing theoretical work~\cite{perez2025radiative, neuman2018origin, groenhof2019tracking, chng2025quantum} employs the Holstein-Tavis-Cummings model (or its multi-mode generalization) to study the exciton-polariton relaxation and utilizes a simple 1D-chain model (depicted in Fig. \ref{fig1}a) to describe the excitonic subsystem. 
 In contrast, most experiments~\cite{Pandya_2022, xu2023ultrafast, mandal2023microscopic, Yongseok2025, polak2020manipulating, pandya2022tuning} use a filled cavity or partially filled cavities with multilayered materials, (as illustrated in Fig. \ref{fig1}b) whose evolution is starkly different, and plays a crucial role in polariton relaxation dynamics. Relative to a single-layer system, a filled cavity houses additional dark excitonic manifolds~\cite{Lydick:24, sun2025exploring, mandal2023microscopic} that open new relaxation pathways (illustrated in Fig. 1) from the upper polariton to the lower polariton. Presently, a microscopic understanding of this polariton relaxation process, especially in filled Fabry-Pérot optical cavities, is missing. In this work, using mixed quantum-classical dynamics, namely the multi-trajectory Ehrenfest approach, as well as an analytical analysis, we provide a microscopic understanding of the polariton relaxation process following an initial excitation in the upper polariton branch of a multilayer material inside an optical cavity.


We show that phonon-induced upper-to-lower polariton relaxation proceeds via two steps. We find that the first step is a phonon-induced interband {\it vertical} transition, i.e., a direct population transfer from the upper polariton to the lower polariton with \textbf{negligible} change in the in-plane polaritonic momentum. The second step is a phonon-induced intraband Fr\"{o}hlich scattering within the lower polariton band. We find that this second step is significantly suppressed in a multilayered (or filled-cavity) geometry due to a {\it phonon-fluctuation synchronization effect}, whereby fluctuations are effectively averaged across layers due to the delocalized nature of the polaritons (via self-averaging). As a consequence of this, the lower polariton population remains energetically localized (i.e., {\it stuck}) for hundreds of femtoseconds. From another perspective, the increase in the number of dark bands appears to suppress phonon-induced Fr\"{o}hlich scattering, protecting the lower polariton from appreciable Fr\"{o}hlich  scattering. We show that the same phonon fluctuation self-averaging plays a central role in polariton relaxation rates, which allows us to obtain simple analytical expressions connecting the number of layers of the material to various relaxation rate constants. Overall, in this work we provide a unified theoretical understanding of exciton-polariton relaxation processes.

\section{Theory}
{\bf  Model.}
To study the relaxation process in a filled cavity, we considered the following light-matter Hamiltonian beyond the long-wavelength approximation~\cite{ mandal2023microscopic, koshkaki2025, keeling2020bose},
\begin{align}
    \hat{H}_{\mathrm{LM}} = \hat{H}_{\mathrm{X}} + \hat{H}_{\mathrm{b}} + \hat{H}_{\mathrm{c}} + \hat{H}_{\mathrm{bX}} + \hat{H}_{\mathrm{cX}} \text{.}
\end{align}
Here $\hat{H}_{X}$, $\hat{H}_{b}$ and $\hat{H}_{c}$ describe the bare exciton, phonon and cavity Hamiltonians. $\hat{H}_{bX}$ and $\hat{H}_{cX}$ describe the interaction between exciton and phonons as well as exciton and the cavity photons, respectively. The excitonic part of the light-matter Hamiltonian is written as
\begin{align}
    \hat{H}_{X} &= \sum^{N,M}_{n,m} \epsilon_0 \hat{X}^\dagger_{n,m} \hat{X}_{n,m} - \tau(\hat{X}^\dagger_{n+1,m} \hat{X}_{n,m} + h.c.)\nonumber\\
    &= \sum^{M}_{m, k} \epsilon_k \hat{X}^\dagger_{k,m} \hat{X}_{k,m}   \text{,}
\end{align}
where $\hat{X}^\dagger_{n,m}$ and $\hat{X}_{n,m}$ are excitonic creation and annihilation operators at site $n$ and layer $m$, $\epsilon_0$ is the on-site energy, and $\tau$ is the intralayer nearest-neighbor hopping integral. Further, $\hat{X}^\dagger_{k,m} = \frac{1}{\sqrt{N}} \sum_{n} e^{ik\cdot x_n} \hat{X}^\dagger_{n,m}$ is the partial Fourier transformed excitonic operator with the in-plane wavevector (momentum) $k$ and with the associated band energy $ \epsilon_k = [\epsilon_0 - 2\tau \cos(k \alpha)]$ where $\alpha$ is the lattice constant. The bare phonon Hamiltonian is written as a collection of harmonic operators of the form
\begin{align}
    \hat{H}_{b} = \sum^N_n \sum^M_m \bigg(\frac{\hat{p}^2_{n,m}}{2} + \frac{1}{2}\omega^2 \hat q^2_{n,m} \bigg) \text{,}
\end{align}
where $\hat{p}_{n,m}$ and $\hat{q}_{n,m}$ are the momentum and position operators of the phonons and $\omega$ is the phonon frequency. The exciton-phonon coupling term is written as
\begin{align}
    \hat{H}_{bX} &= \gamma \sum^N_n \sum^M_m \hat q_{n,m} \hat{X}^\dagger_{n,m} \hat{X}_{n,m} \nonumber \\
    &= \frac{\gamma}{{N}} \sum_{k, k'} \sum^M_m  \hat{X}^\dagger_{k,m} \hat{X}_{k',m}  \sum_n \hat{q}_{n,m} e^{i(k-k')x_n}\label{eq:phonon-coupling}
\end{align}
where $\gamma$ is the exciton-phonon coupling strength.
The cavity part of the light-matter Hamiltonian considers a set of confined radiation modes (with differing in-plane momentum) in a Fabry-Pérot optical cavity  and is written as (using $\hbar = 1$ a.u.)
\begin{align}
    \hat{H}_{c} = \sum_k \omega_k \hat{a}^\dagger_k \hat{a}_k \text{,}
\end{align}
where $\hat{a}^\dagger_k$ and $\hat{a}_k$ are creation and annihilation operators for wavevector $k$. The photon frequency is described by
\begin{align}
    \omega_k = \frac{c}{\eta}\abs{k \Vec{x} + k_0 \Vec{y}} \text{,}
\end{align}
where $c$ is the speed of light, $\eta$ is the refractive index, and $\vec{x}$ and $\vec{y}$ are unit vectors along $x$ and $y$ directions. Further, the photon wavevector in the quantization direction ($y$-direction) is $k_0 = \frac{\pi}{L}$, where $L = 1000 ~\mathrm{\AA}$ is the distance between the two cavity mirrors. The in-plane momentum is $k = \frac{2 \pi n_x}{N \cdot \alpha}$ where we have imposed a periodic boundary condition in  the lateral direction with $n_x$ = 0, $\pm$ 1, $\pm$ 2, and $\alpha$ = 12 \r{A}. 
Finally, we consider the exciton-cavity interaction $\hat{H}_{cX}$ beyond the long-wavelength approximation which is written as
\begin{align}
   \hat{H}_{cX} = \sum_{n, m, k} \frac{\Omega_k}{\sqrt{N}} \bigg(\hat{X}^\dagger_{n, m} \hat{a}_k e^{ikx_n} + h.c.\bigg) \sin(k_0\cdot y_m) \text{,}
\end{align}

\begin{figure*}
\centering
\includegraphics[width=1.0\linewidth]{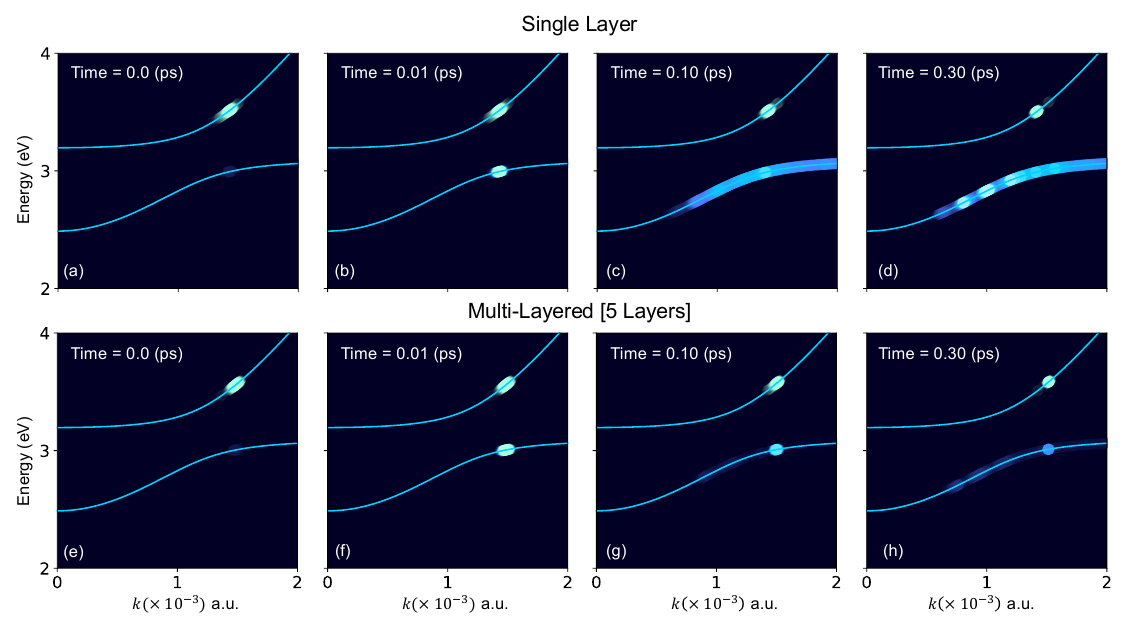}
\caption{\footnotesize \textbf{Polariton-band-resolved relaxation dynamics.} Polariton-band-resolved population relaxation dynamics in a single-layer material ({a-d}), and in multi-layered material (5 layers) ({e-h}), following an excitation to the upper polariton centered at 3.5 eV. The plots from left to right display a snapshot at the initial time, 0.01 ps, 0.10 ps, and 0.30 ps of the population (blue circles), where there is not only a transfer of population from the upper to lower polariton, but also Fr\"{o}hlich scattering the lower polariton. The degree of this scattering (a-d) is much greater for a single layer of material, than in the multilayered material (e-h), owing to the "phonon-fluctuation synchronization effect" in the multilayered case.}
\label{fig2}
\end{figure*}

where $\Omega_k = \sqrt{\frac{\omega_0}{\omega_k}} \Omega_0$ is the light matter coupling strength. In this work, the exciton-photon subsystem is treated quantum mechanically while the phonon degrees of freedoms are evolved quasi-classically using the multi-trajectory Ehrenfest approach. We restrict our dynamics to the single excitation subspace such that the wavefunction describing the exciton-photon subspace is written as
\begin{align}
    \ket{\Psi(t)} =  \sum_k c_k(t) \ket{1_k} + \sum_{n,m} b_{n,m}(t) \ket{n,m} \text{,}
\end{align}
with $c_k(t)$ and $b_{n,m}(t)$ are the time-dependent coefficients, $\ket{1_k} \equiv \hat{a}^\dagger_k \ket{\bar{0}}$ and $\ket{n,m} \equiv \hat{X}^\dagger_{n,m} \ket{\bar{0}}$ with $\ket{\bar{0}}$ representing the ground state (or vacuum). 

The bare exciton-polariton Hamiltonian $\hat{H}_{EP} = \hat{H}_{LM} - \hat{H}_{b} - \hat{H}_{bX}$ can be re-expressed as,
\begin{align}
    \hat{H}_{EP} = \sum_{k, i \in \pm} \omega_{k,i} \hat{P}^\dagger_{k,i} \hat{P}_{k,i} + \sum_{k,d} \epsilon_k  \hat{X}^\dagger_{k,d} \hat{X}_{k,d} \text{,}
\end{align}

where $\{\hat{P}^\dagger_{k,\pm}\}$  are the upper ($+$) and lower ($-$) polariton operators and $\{\hat{X}^\dagger_{k,d}\}$
are dark excitonic operators. The polaritonic operators are defined as 
\begin{align}
    \hat{P}^\dagger_{k,+}=\sin(\theta_k) \hat{a}^\dagger_k + \cos(\theta_k) \hat{X}^\dagger_{k,B}  \text{,}\\
    \hat{P}^\dagger_{k,-} = \cos(\theta_k) \hat{a}^\dagger_k - \sin(\theta_k) \hat{X}^\dagger_{k,B} ,
\end{align}
where $\hat{X}^\dagger_{k,B} = \frac{1}{\sqrt{S}}\sum_{m}^{N_L} \sin(k_0\cdot y_m) \hat{X}^\dagger_{k,m} $ is the bright exciton creation operator that creates an exciton delocalized in all $N_L$ layers with in-plane momentum $k$ and $\theta_k = \frac{1}{2} \tan^{-1}[2\sqrt{S}\Omega_k/(\omega_k-\epsilon_k)]$ is the mixing angle with $S = \sum_m \sin^2(k_0 \cdot y_m)$. Further, $\hat{X}^\dagger_{k,d} = \sum_{m}\mathcal{D}_{m,d} \hat{X}^\dagger_{k,m}$ where $\mathcal{D}_{m,d}$ are orthonormal coefficients ($\sum_{m,m'}\mathcal{D}_{m,d}\mathcal{D}_{m',d'} = \delta_{m,m'} \delta_{d,d'}$) which satisfy $\sum_m \mathcal{D}_{m,d} \cdot \sin(k_0 \cdot y_m)  = 0$, for all $d$, such that $[\hat{X}^\dagger_{k,d}, \hat{X}_{k,B}] = 0$. \\

{\bf  Quantum dynamical approach.} Due to the formidable nature of exact quantum dynamical simulations, here, we use the multi-trajectory Ehrenfest approach (as in recent works~\cite{chng2025quantum, ying2025microscopic, groenhof2019tracking, xu2023ultrafast, blackham2025microscopic, tichauer2021multi, hoffmann2020effect, hoffmann2019capturing}) to propagate the quantum dynamics of the light-matter system. In this mixed quantum-classical approach, the phonons are propagated quasi-classically using the following equations of motion:
\begin{align}
    \dot{p}_{n,m}(t) = - \bra{\Psi(t)} \frac{\partial \hat{H}_{LM}}{\partial q_{n,m}} \ket{\Psi(t)} \text{,} \hspace{0.25cm} \dot{q}_{n,m}(t) = p_{n,m}(t) \text{.}
\end{align}
The exciton-polariton part is propagated quantum mechanically using an efficient split-operator approach combined with a bright-layer unitary transformation~\cite{koshkaki2025}. Short-time propagation of the exciton-polariton wavefunction over a time step $\delta t$ is written as
\begin{align}
    \ket{\Psi (t+\delta t)} = \hat{U}_{pol} e^{-i \hat{H}_{EP} \delta t} \hat{U}^\dagger_{pol} e^{-i \hat{H}_{env} \delta t}  \ket{\Psi (t)} \text{,}
\end{align}
where $\hat{U}^\dagger_{pol}$ denotes the unitary transformation that maps the real-reciprocal space basis to the polaritonic basis, and $\hat{H}_{env}= \hat{H}_b + \hat{H}_{bX}$, which represents the phononic contribution to the Hamiltonian. Furthermore, both $\hat{H}_{EP}$ and $\hat{H}_{env}$ are each diagonal in their own natural bases; their combination can be evaluated through an element-wise operation rather than a full matrix multiplication, leading to a substantial reduction in computational cost. Our full mixed quantum-classical dynamics method is described in Ref.~\cite{koshkaki2025}. 

\begin{figure*}
\centering
\includegraphics[width=1.0\linewidth]{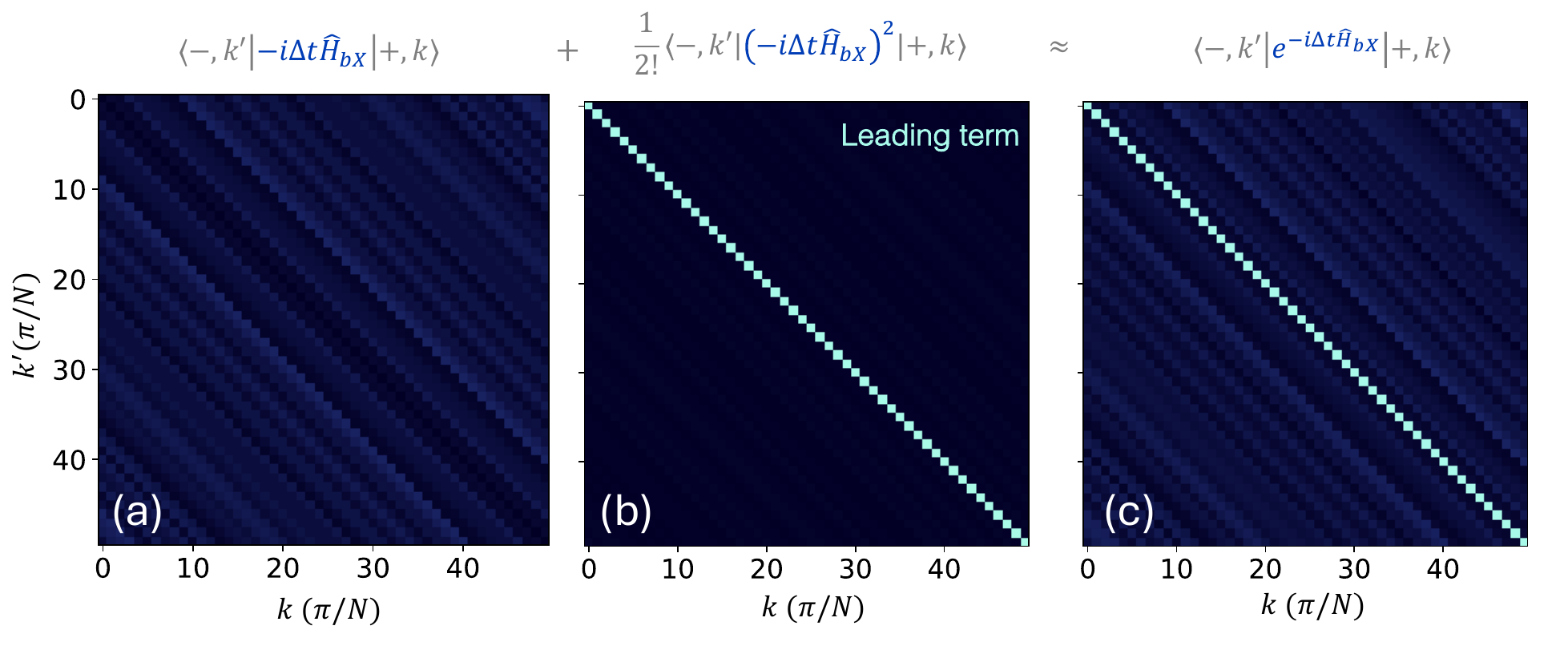}
\caption{\footnotesize \textbf{The vertical nature of the polariton relaxation.} ({a - c}) Upper to lower polariton transfer matrix elements for the first-order (a), second-order (b), and full exponential propagator (c). Note that here we are presenting the absolute values of the matrix elements. Here, we consider a system comprising 40,000 sites and a time step of $\Delta t = 50$ a.u.}
\label{fig3}
\end{figure*}

{\bf  Parameters.} For the  simulations            presented below, the on-site excitonic energy is set to $\epsilon_0 = 3.2~\mathrm{eV}$, and the nearest-neighbor hopping parameter is chosen as $\tau = 400~\mathrm{cm^{-1}}$. The intralayer lattice constant is set to $\alpha = 12~\mathrm{\AA}$, and an interlayer spacing is set to $\alpha_y = 40~\mathrm{\AA}$. The two mirrors are separated by a distance $L = 1000~\mathrm{\AA}$, and light-matter coupling is $\Omega_0 = 241.7$ meV for a single-layer material. For multilayered materials we re-normalize $\Omega_0$ by $1/\sqrt{\mathcal{S}}$ to make a fair comparison wherein the overall Rabi splitting is the same regardless of the thickness of the material. The phonon frequency is $\omega = 720~\mathrm{cm^{-1}}$ and the exciton-phonon coupling strength is $\gamma = 3.76 \times 10^{-4}$ a.u.  


{\bf Extracting Rate Constants.} The relaxation dynamics among the upper, lower, and dark polaritonic states can be modeled as a three-state kinetic model (coupled rate equations) involving a set of rate constants which is written as
\begin{align}
\frac{d}{dt}\Vec{P} = \mathbf{K}\Vec{P},
\end{align}
where $\Vec{P} = [P_{+}(t), P_{d}(t), P_{-}(t)]^{T}$ is the population vector, which includes the upper polariton, dark excitons, and lower polariton populations, respectively, where $T$ denotes a matrix transpose.  These quantities are computed as

\begin{align}
P_{+}(t) &= \sum_k \bra{\bar{0}} \hat{P}_{k,+} \ket{\psi(t)}, \\
P_{-}(t) &= \sum_k \bra{\bar{0}} \hat{P}_{k,-} \ket{\psi(t)},\\
P_{d}(t) &= 1 - P_{+}(t) - P_{-}(t).
\end{align}
The rate constant matrix $\mathbf{K}$ describing the interconversion among these states is expressed as  
\begin{align}
\mathbf{K} =
\begin{bmatrix}
       -k_{UL} - k_{UD} & k_{DU} & k_{LU} \\
       k_{UD} & -k_{DL} - k_{DU} & k_{LD} \\
       k_{UL} & k_{DL} & -k_{LU} - k_{LD}
\end{bmatrix},
\end{align}
where each $k_{ij}$ denotes the transition rate constant from state $i$ to $j$ (e.g., $k_{UL}$: upper $\rightarrow$ lower, $k_{UD}$: upper $\rightarrow$ dark, $k_{LD}$: lower $\rightarrow$ dark), as schematically illustrated in Fig. \ref{fig1}.  
This system of coupled first-order linear differential equations can be analytically solved via eigenvalue decomposition, yielding the expression given as
\begin{align}
\Vec{P}(t) = e^{\mathbf{K}t}\Vec{P}(0).
\end{align}
The rate constants are extracted by fitting the analytical model to the computed population dynamics.

 \section{Results}
Fig. \ref{fig2}  presents the relaxation dynamics in single-layer material and in a multilayered material (or material of a finite thickness) inside an optical cavity.  Here the system is initially excited within an energy window of $3.5 \pm 0.2$ eV in the upper polariton. The initial polariton-band-resolved population is illustrated in Fig. \ref{fig2}a and Fig. \ref{fig2}e for the single-layer and five-layer configurations, respectively.

Fig. \ref{fig2}b, and f shows the lower polariton branch population after propagation for $0.01$ ps, and this relaxation process can be seen to approximately preserve polaritonic momentum leading to a  vertical ($k \rightarrow k$) transition. While a phonon-induced relaxation that enables fast population transfer between polaritonic branches is expected, the vertical nature of the transition is counterintuitive. In the following (Fig. \ref{fig3}) we provide a microscopic mechanistic understanding of this  vertical nature of the polariton relaxation. 

Fig. \ref{fig2}c-d shows that, at longer times, the population within the lower polariton for a single-layered material is highly scattered owing to the phonon fluctuations. Interestingly, in the multilayered material, this phonon induced scattering is significantly suppressed, leading to a relaxation process that continues to conserve momentum for a longer duration (see Fig. \ref{fig2}g-h), which we attribute to a phonon-fluctuation synchronization effect that reduces the magnitude of the phonon fluctuations, suppressing the Fr\"{o}hlich scattering.


Our numerical results demonstrate that the first step of the polaritonic relaxation is (polariton) momentum conserving. This is surprising, as the phonon-induced relaxation that originates from $\hat{H}_{bX}$ does not directly show an in-plane momentum conservation, as a transition from  $k \rightarrow k'$  appear to be allowed. To clearly see this, consider the second line of Eq.~\ref{eq:phonon-coupling} which can be expanded (replace $\{\hat X_{n,m}^{\dagger}\}$ with $\{\hat{P}_{k,i}^{\dagger}, \hat X_{k,d}^{\dagger}\}$ ) to obtain the following polariton-phonon coupling term responsible for the upper-to-lower polariton relaxation:
\begin{equation}\label{P2P}
 \mathds{P}_{-} \hat{H}_{bX}  \mathds{P}_{+} = \gamma \sum_{k,k'} \hat{P}_{k,-}^{\dagger}\hat{P}_{k',+} \Big[\cos{(\theta_k)}\sin(\theta_{k'}) \mathcal{Q}(t)\Big].
\end{equation}

Here, $\mathds{P}_{\pm} = \sum_{k} \hat{P}_{k,\pm}^{\dagger}\hat{P}_{k,\pm}$ are projection operators and $\mathcal{Q}(t) = \sum_{n} e^{-i(k-k')x_n} \frac{1}{S}\sum_m q_{n,m} \sin^2{(k_0 \cdot y_m)}$ is a time-dependent variable that encodes the dynamical fluctuations of the phonons. Clearly Eq.~\ref{P2P} suggests that $\hat{P}_{k,+}^{\dagger} \rightarrow \hat{P}_{k',+}$ transitions are allowed for $k\neq k'$ which contradicts the results presented in  Fig. \ref{fig2}a-b and Fig. \ref{fig2}e-f. Below we resolve this apparent contradiction.

{\bf  Polaritonic vertical transition.} To find the microscopic mechanism of the (predominantly) vertical nature of the polaritonic transition at early times, we compute the transition probability to the lower polariton, following an excitation to the upper polariton, after a short time propagation of  $\Delta t$, which is computed as
\begin{align}
P_{-,k'}( \Delta t)
&\approx \left| \langle -,k' | e^{-i\hat{H}_{bX} \Delta t} e^{-i\hat{H}_{EP} \Delta t} | +,k \rangle \right|^2 \nonumber \\
&=  \left| \langle -,k' | e^{-i\hat{H}_{bX} t} | +,k \rangle \right|^2,
\end{align}
where $\ket{+,k} = \hat{P}^\dagger_{k,+}\ket{\bar{0}}$ and $\ket{-,k} = \hat{P}^\dagger_{k,-}\ket{\bar{0}}$. Note that we arrive at the second line by using the fact that $e^{-i\hat{H}_{EP} \Delta t} |+,k\rangle = e^{-iE_{+,k} \Delta t} |+,k\rangle$ since the polaritonic states are the eigenstates of the bare exciton-polariton Hamiltonian $\hat{H}_{0}$.
\begin{figure*}
\centering
\includegraphics[width=1.0\linewidth]{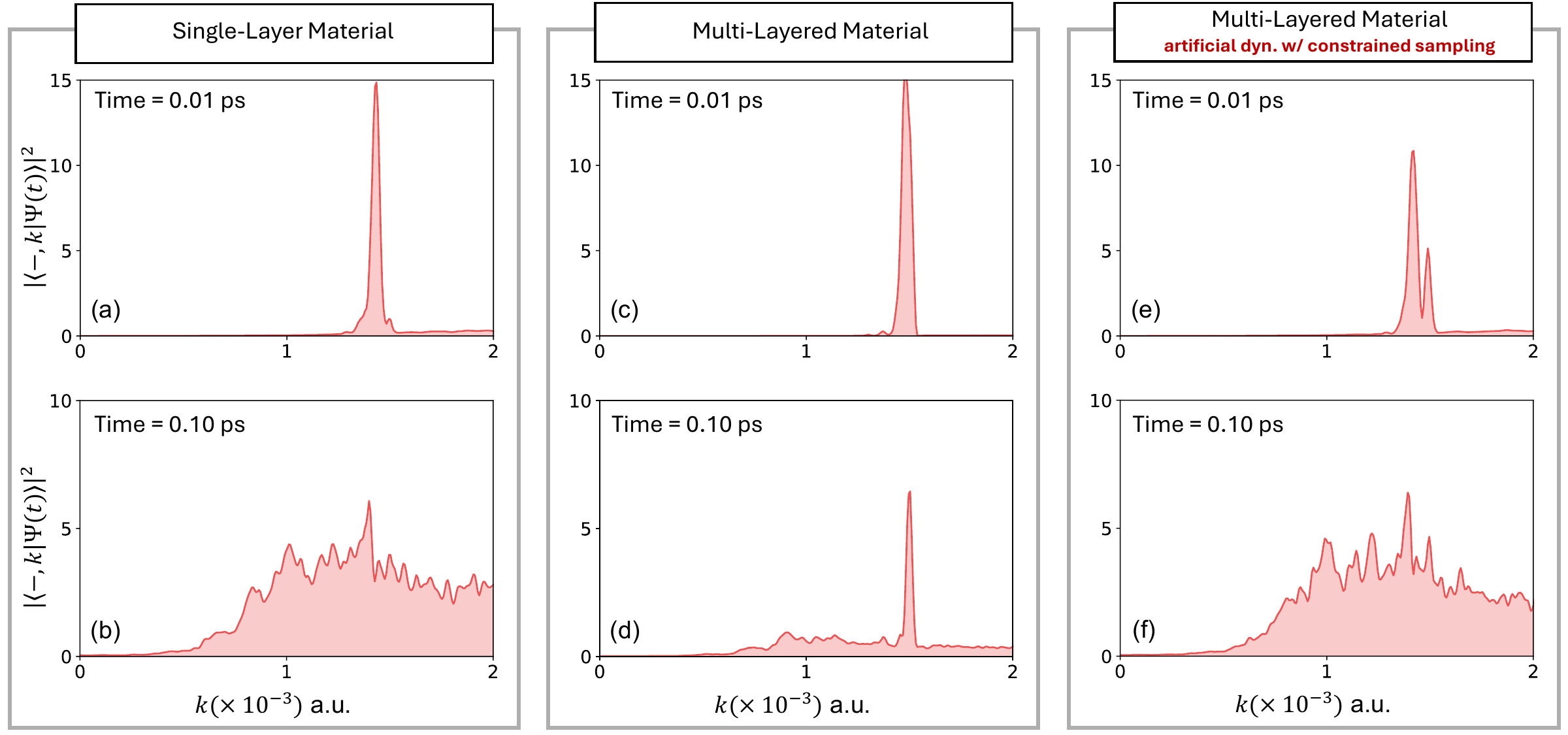}
\caption{\footnotesize \textbf{Suppression of Fröhlich scattering in  multilayered materials via phonon fluctuation synchronization effect.} ({a - d}) Show the relaxation dynamics in the lower polariton band in single layer vs multilayered material. ({e - f}) Show the relaxation dynamics in the lower polariton when we constrain phonon sampling. For this simulation we copied the initial position and momentum of the phonons for each layer, eliminating any difference in the phonons layer to layer.}
\label{fig4}
\end{figure*}
Fig.~\ref{fig3}c presents $\left| \langle -,k' | e^{-i\hat{H}_{bX} t} | +,k \rangle \right|$, which clearly demonstrates that the diagonal transition elements dominate, leading to vertical transitions. To investigate the origin of this behavior, we consider the Taylor expansion of $e^{-i\hat{H}_{bX} \Delta t}$, written as
\begin{align}
    e^{-i\hat{H}_{bX}\Delta t } = \mathds{1} -i\hat{H}_{bX}\Delta t + \frac{(i\hat{H}_{bX}\Delta t)^2}{2!} + ...
\end{align}
The term $\mathds{1}$ will not contribute to the upper-to-lower polariton transition and can be dropped. Interestingly, we find that the contribution of the term linear in $\Delta t$  is also vanishingly small (see Fig.~\ref{fig3}a) while the leading non-negligible contribution comes from the second order term in $\Delta t$, (see Fig.~\ref{fig3}b) with its diagonal elements dominating.  




This behavior can be analytically understood by noting that $(\sum_n\hat{X}^\dagger_n\hat{X}_nR_n)^p = \sum_n \hat{X}^\dagger_n\hat{X}_n R_n^p$, where $p \in \{1, 2, ..., N\}$, allowing for the transition element to be expressed as

\begin{align}
    \langle -,k' &| e^{-i\hat{H}_{bX} \Delta t} | +,k \rangle \nonumber  \\&=   C_{kk'}  \sum_{n} \frac{e^{i(k'-k)\cdot x_n}}{N} (1 - e^{-i\gamma q_{n,m}\Delta t} )\nonumber \\
    &\approx    C_{kk'} 
    \langle  1 - e^{-i\gamma q\Delta t} \rangle  \sum_{n} \frac{e^{i(k'-k)\cdot x_n}}{N}\nonumber \\
   & =  C_{kk'} 
    \langle  1 - e^{-i\gamma q\Delta t} \rangle  \cdot \delta_{kk'} 
\end{align}
where $C_{kk'}= \sum_m \sin\theta_{k'} \cos\theta_k \frac{\sin^2(k_0 \cdot y_m)}{\mathcal{S}} $ and $\langle ... \rangle$ represents phase space averaging. Importantly, here we arrive at the third line by utilizing the fact that for the relevant dynamics, $k-k' \ll 1/\alpha$ which is valid for $\alpha = 12 ~\mathrm{\AA}$ set here and in typical materials. Using the fact that typical relaxation process occur between the anti-crossing points, the following criteria can be derived 
\begin{align}\label{criteria}
 \frac{\eta}{c} \sqrt{\epsilon_0^2 - \omega_0^2 } \ll \frac{1}{\alpha}
\end{align}
for which a material will exhibit vertical transition. It is worth noting that artificially scaling lattice constant (as is done in  recent work~\cite{krupp2025quantum}) will miss the vertical nature of this polaritonic relaxation as such model will not satisfy the criteria in Eq.~\ref{criteria}. Note that the magnitude of the transition is governed by the phonon disorder term
\begin{align}
    \left\langle 1- e^{-i\gamma q \Delta t } \right\rangle 
&= 
i \gamma  \Delta t \left\langle q \right\rangle
+ \frac{(\gamma  \Delta t)^2}{2} \left\langle q^2 \right\rangle
+ \cdots
\end{align}
Here, the first-order term is negligible as $\braket{q} \approx 0$ for harmonic phonon modes. In contrast, the second-order term  $\braket{{q}^2 } = [2\omega \tanh(\beta\omega/2)]^{-1} \approx \frac{1}{\beta\omega^2}$ (with the approximated form obtained in the classical limit) is the leading (non-negligible) term which corroborates our findings in Fig.~\ref{fig3}.    

As a result of this vertical transition, we observe an energetically localized  density in the lower polariton at short times (see Fig.~\ref{fig2}b and Fig.~\ref{fig2}f). In the single-layer material, at longer times, this energetically localized population density spreads out due to phonon-induced Fr\"{o}hlich scattering. However, in the multilayered material (or material of finite thickness), this spreading is significantly suppressed.  Below we show that this is due to a phonon fluctuation synchronization effect, where phonon fluctuations are averaged over multiple layers.

{\bf Fr\"{o}hlich scattering.} To clearly understand the suppression of the Fr\"{o}hlich scattering in the lower-polariton, we consider the phonon interaction term projected within the bright excitonic subspace that is written as

\begin{align}\label{exc-phonon2}
\mathds{P}_{B}   \hat H_\mathrm{bX}  \mathds{P}_{B}   &= \gamma  \sum_{n} \hat{X}_{n,B}^{\dagger}\hat{X}_{n,B} \bar{q}_{n}. 
\end{align} 
where $\bar{q}_{n} = \frac{1}{S}\sum_m  \sin^2(k_0 \cdot y_m) q_{n,m}(t)$ is layer-averaged phonon fluctuation and $\mathds{P}_{B} = \sum_n \hat{X}_{n,B}^{\dagger}\hat{X}_{n,B}$. Because the bright exciton couples only to this weighted average over layers (i.e., $\bar{q}_{n}$), local phonon fluctuations are partially canceled in the sum, as they are uncoupled and random, which leads to the effective suppression of the phonon-induced disorder. We refer to this as the phonon-fluctuation synchronization effect. In our recent work~\cite{koshkaki2025}, we have shown that this phonon-fluctuation synchronization effect enables enhanced coherent transport and significantly increases coherence lifetime of the polariton. Note the arguments based on the bright layer projected phonon couplings in Eq.~\ref{exc-phonon2} are valid when the dark layers are energetically decoupled due to the light-matter couplings with upper and lower polaritons lying above and below the dark-excitonic energy. 
\begin{figure}
\centering
\includegraphics[width=1.0\linewidth]{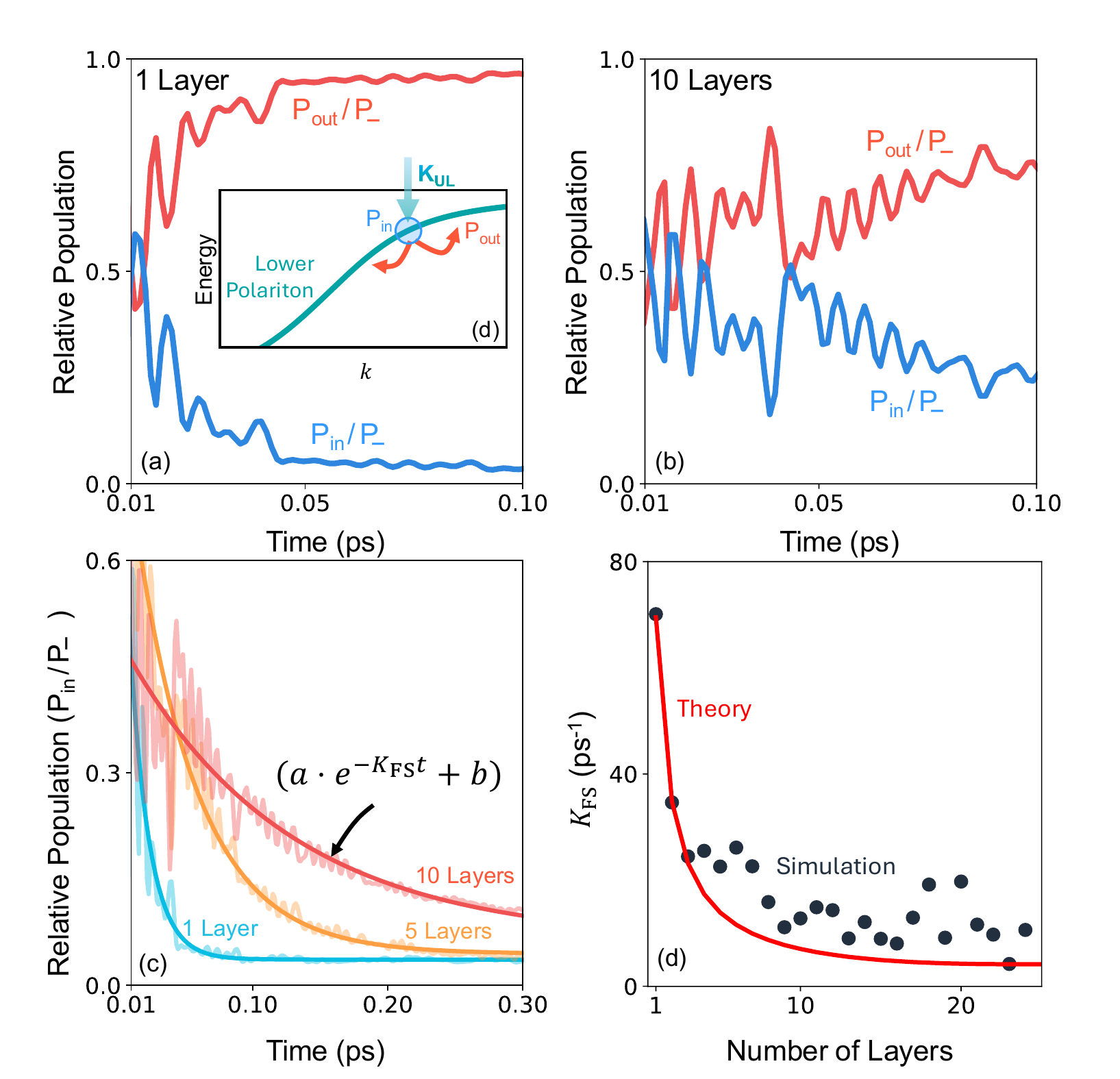}
\caption{\footnotesize \textbf{Population  inside and outside the excitation window.} ({a \& b}) Population, relative to the lower polariton, inside the coherent relaxation window vs population scattering outside this window in a single layer and in 10 layers. ({c}) Inside population and fitted exponential function for 1, 5, and 10 layers. ({d}) Fitted $K_{FS}$ compared to our analytical expression as a function of the number of layers.}
\label{fig5}
\end{figure} 
The extent of the suppression of phonon fluctuations can be quantified by the effective phonon variance, defined as
\begin{align}\label{q2} \langle \bar{q}^2 \rangle = \frac{1}{N}\sum_{n} \bar{q}_n^2 = \langle {q}^2 \rangle \cdot \frac{1}{S^2}\sum_m^{N_L} \sin^4(k_0 \cdot y_m) \end{align}

where the factor $\frac{1}{S^2}\sum_m  \sin^4(k_0 \cdot y_m) \le 1$ encodes the reduction due to multilayer averaging of the phonon fluctuations. The phonon fluctuation decreases with an increase in the number of layers which also increases the number of dark layers. Thus, interestingly the presence of  a large number of dark layers appear to (indirectly) sequester phonon-induced fluctuation leading to suppressed Fr\"{o}hlich scattering. At the same time, $\langle {q}^2 \rangle \approx \frac{1}{\beta\omega^2}$, such that this fluctuation synchronization can also be viewed as a suppression of the effective classical temperature which is expressed as
\begin{align} 
T_{\mathrm{eff}} = T \cdot \frac{1}{S^2}\sum_m \sin^4(k_0 \cdot y_m).
\end{align}

\begin{figure*}
\centering
\includegraphics[width=1.0\linewidth]{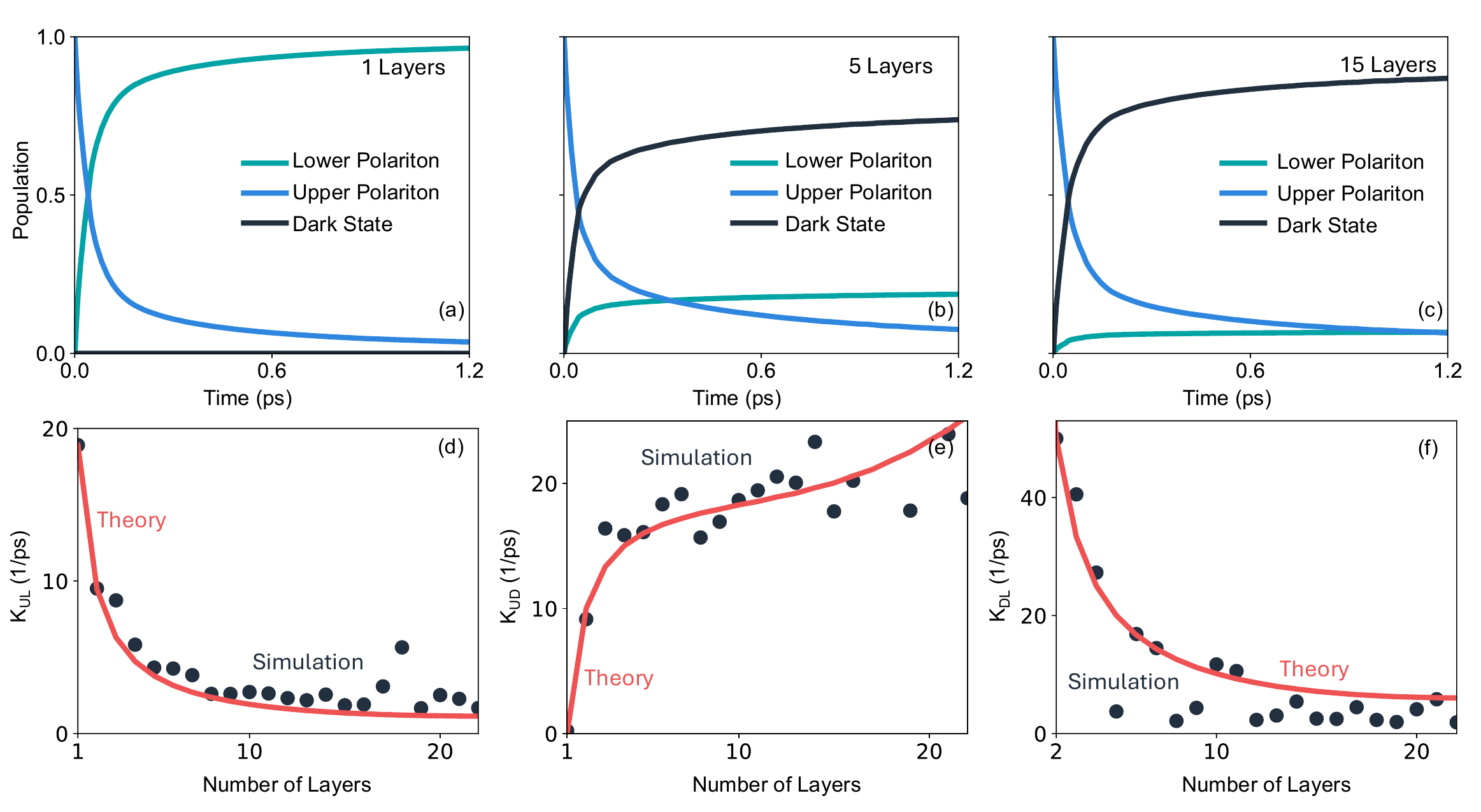}
\caption{\footnotesize \textbf{Effect of number of layers on polariton dynamics and kinetics.} ({a-c}) Population dynamics for 1, 5 and 15 Layers, initialized at 3.5 eV and using $\omega = 720cm^{-1}$, $\tau = 400 cm^{-1}$. ({d-f}) Relaxation rate from upper polariton to lower polariton $K_{UL}$ ({d}), upper to dark $K_{UD}$ ({e}), and dark to lower $K_{DL}$ ({f}), vs number of layers for excitation in the upper polariton at 3.5 eV. }
\label{fig6}
\end{figure*}
In Fig.~\ref{fig4}, we numerically test our proposed mechanism: Fr\"{o}hlich scattering is suppressed by the phonon-fluctuation synchronization effect, leading to a long-lived, $k$-space–localized population in multilayered materials. Fig.~\ref{fig4}a-b  and Fig.~\ref{fig4}c-d presents the k-resolved lower polariton population (with the system initialized in the upper polariton) in a single-layer material and a multi-layered material, respectively. As was shown in Fig.~\ref{fig2}, while the initial density is k-localized at short times in the lower polariton for both cases ($\sim10$fs), at longer times the population density remains $k$-localized only in the multilayered material. To provide a direct numerical evidence of the phonon synchronization mechanism, in Fig.~\ref{fig4}e-f we perform a quantum dynamical simulation where we sample the phonons in a constrained fashion. We set $q_{n,m}(0) = q_{n,1}(0)$, for all layers, such that $\bar{q}_n(0) =  q_{n,1}(0)$. Under this constrained initial sampling, the weighted average of identical fluctuations over multiple layers is the same as the fluctuations themselves. Under such constrained sampling, our numerical simulation, presented in Fig.~\ref{fig4}e-f, show that the dynamics resembles the single layer scenario where the $k$-localized lower polariton density at short times (see Fig.~\ref{fig4}e) become  $k$-delocalized in the same fashion (see Fig.~\ref{fig4}f)  as in the single layer scenario. This numerically demonstrates that the $k$-localization in the multi-layered scenario originates from the phonon fluctuation averaging.

We find that the Fr\"{o}hlich scattering rate is directly related to the variance of the phonon fluctuations $\langle \bar{q}^2 \rangle$, as described within rate theories for time-dependent Peierls coupling~\cite{ChowdhuryJCP2021, troisi2003rate, beratan2009steering}, since scattering proceeds through the time-dependent coordinates $q(t)$. To clearly see this, consider the multi-layer dependence of the Fr\"{o}hlich scattering rate $K_{FS}$  which can be obtained by analyzing the time-dependent couplings $V_{FS}(t) = \langle -, k| \hat{H}_{\mathrm{bX}}(t)|-, k'\rangle  \propto \langle -, k|\mathds{P}_{B} \hat{H}_{\mathrm{bX}}(t)\mathds{P}_{B}|-,k'\rangle $ where $\mathds{P}_{-}$ is a lower polariton projection operator. For such a time-dependent coupling term $V_{FS}(t)$ the rate constant is known~\cite{ChowdhuryJCP2021, troisi2003rate, beratan2009steering} to be proportional to the variance $\langle V_{FS}^2(t)\rangle$. Therefore the Fr\"{o}hlich scattering rate constant $K_{FS}$ is directly proportional to $\langle \bar{q}^2 \rangle$ (following the steps in Eq.~\ref{exc-phonon2} - Eq.~\ref{q2}) and is then expressed as
\begin{align}\label{KFS}
K_{FS} (N_L) = A_{{FS}} \cdot  \frac{1}{S^2}\sum_m^{N_L} \sin^4(k_0 \cdot y_m)
\end{align}
 where $N_L$ is the number of layers and $A$ is a prefactor. The prefactor can also be written as $A_{FS}  = K_{FS}(1)$ which corresponds to the $K_{FS}$ in the single-layer material setup which we use in our analytical expression. 

 In Fig.~\ref{fig5} we numerically analyze the Fr\"{o}hlich scattering rate in the lower polariton for the data presented in Fig.~\ref{fig2} and Fig.~\ref{fig4}. We compute the relative lower polariton population inside and outside the $k$-window ($\bar k \pm 5\cdot(2\pi/N\alpha)$ with the initial excitation centered around $\bar k$) within which the system was excited in the upper polariton. Fig.~\ref{fig5}a-b show that the relative inside population $P_{in}/P_{-}$ decays less significantly for 10 layers compared to a single layers. We fit this relative inside population (see Fig. \ref{fig5}c)  to an exponential curve (i.e., $a \cdot e^{-K_{FS}t} + b$ where $a$ and $b$ are scalar constants) to estimate the rate ($K_{FS}$) of Fr\"{o}hlich scattering induced by the phonons. 
 
 Fig.~\ref{fig5}d  presents the layer dependent Fr\"{o}hlich scattering rate computed numerically compared to the analytical scaling expression provided in Eq.~\ref{KFS}. Overall, our results demonstrate that the qualitative layer-dependence of Fr\"{o}hlich scattering rate is very well captured by our theoretical expression validating the microscopic mechanisms revealed in this work. 

{\bf Polariton relaxation rates.} Fig.~\ref{fig6} presents the how population dynamics is modified when altering the thickness of the material (by altering the number of layers). Fig.~\ref{fig6}a-c show that increasing the number of layers drastically reduces the lower polariton population due to the competition from the dark states that are excitonic. We also observe this in the upper-to-lower polariton rate constant $K_{UL}$ that decreases with increase in number of layers. Following the same strategy as in the case of Fr\"{o}hlich scattering rate constant, we arrive at the following analytical expression 
\begin{align}\label{KUL}
K_{UL} (N_L) = A_{UL} \cdot  \frac{1}{S^2}\sum_m^{N_L} \sin^4(k_0 \cdot y_m).
\end{align}

Just as before we use $A_{UL} = K_{UL} (1)$ and obtain the theoretical curve presented in Fig.~\ref{fig6}d. Our analytical theory  accurately captures the numerically extracted upper-to-lower rate constant.

The same strategy can again be utilized to provide an analytical expression for the upper-to-dark relaxation rate constant. Here, the relevant time-dependent coupling term between the upper polariton to the $d$th dark layer is  $V_{UD}^{d}(t) = \langle +, k| \hat{H}_{\mathrm{bX}}(t)|k,d \rangle$  where $|k,d \rangle = \hat{X}^{\dagger}_{k,d} |\bar0\rangle$. Using this, we find that the overall upper-to-dark relaxation rate constant (which sums over all dark layers) can be  written as
\begin{align}\label{KUD}
K_{UD} (N_L) &= A_{UD}  \sum_{m}\frac{\sin^2(k_0\cdot y_m)}{{S}}\sum_{d}{\mathcal{D}_{m,d}^2} \nonumber \\
&\approx  A_{UD} (N_{L}-1) \cdot\sum_{m}\frac{\sin^4(k_0\cdot y_m)}{{S}^2} 
\end{align}
where $A_{UD}$ is a prefactor (which we obtain via fitting), and we have made the approximate replacement $\sum_{d}{\mathcal{D}_{m,d}^2} \approx (N_{L}-1) {\sin^2(k_0\cdot y_m)}/S$ to provide a simple analytical expression. As a result of the additional $(N_L-1)$, the upper-to-dark polariton relaxation rate constant $K_{UD}$ rapidly increases with the number of layers and then saturates at a certain number of layers. Note that the late rise in the rate constant (see red solid line in Fig.~\ref{fig6}e) after $\sim$15 layers found in our theory vanishes when not making the approximation in the second line in Eq.~\ref{KUD}.

Similarly, the dark-to–lower polariton relaxation rate can be described within the same analytical framework. The relevant time-dependent coupling between the lower polariton and the 
$d$-th dark layer is given by $V_{DL}^{d}(t) = \langle k, d| \hat{H}_{\mathrm{bX}}(t)|-,k \rangle$,  where $|k,d \rangle = \hat{X}^{\dagger}_{k,d} |\bar0\rangle$. Using this coupling, the total dark-to–lower polariton relaxation rate constant can be expressed as follows
\begin{align}
    K_{DL} (N_L) &= A_{DL}  \sum_{m}\frac{\sin^2(k_0\cdot y_m)}{{S}}\sum_{d} {\mathcal{D}_{m,d}^2} \nonumber \\
&\approx  A_{DL}  \sum_{m}\frac{\sin^4(k_0\cdot y_m)}{{S}^2} .
\end{align}
Here, $A_{DL} = K_{DL}(1)$, from which we obtain the theoretical curve shown in Fig.~\ref{fig6}f. The analytical model accurately reproduces the numerically extracted dark-to-lower rate constant. The factor $N_L - 1$ does not appear because the rate equation $\sum_{d}^{N-1} \frac{dP_d(t)}{dt} \approx -\sum_{d}^{N-1} K_{DL} P_d(t)$ is summed over 
$d$ on both sides. Note that while the lower polariton can also be populated via upper-to-dark and then dark-to-lower polariton relaxation, such a two-step kinetic process plays a relatively insignificant role when compared to a direct upper-to-lower polariton relaxation process for the range of timescales (picoseconds) presented in this work.  Overall, the phonon fluctuation synchronization effect successfully explains various relaxation processes, specifically how they are modified in a material of finite thickness placed inside an optical cavity.

\section{Summary and Conclusion} 

In conclusion, we have explored exciton-polariton relaxation dynamics following an excitation in the upper polariton branch. Using direct mixed quantum-classical quantum dynamical simulations as well as analytical analysis, we provide the mechanistic principles that govern exciton-polariton relaxation and show how the finite thickness of materials modifies these processes.  

Our results demonstrate that the polariton relaxation proceeds through a two-step mechanism involving an interband vertical transition followed by intraband phonon-mediated Fröhlich scattering. We provide an analytical criterion for the molecular lattice constant; typical molecular parameters satisfy this criterion, enabling such vertical transitions.

Furthermore, we find that the second step, namely the Fröhlich scattering within the lower polariton band, is significantly suppressed when the finite thickness of a material is taken into consideration. We attribute this reduction to the phonon-fluctuation synchronization effect, where the spatial delocalization of polaritons induces self-averaging of phonon fluctuations across layers along the quantization axis, effectively weakening Fröhlich scattering in the lower polariton. As a result, cavities hosting materials of finite thickness (i.e., multilayered materials) exhibit a long-lived,  $k$-localized polaritonic density upon polariton relaxation. We show that an analytical description of various relaxation processes can be developed based on the phonon-fluctuation self-averaging mechanism. We find that increasing the number of layers suppresses both upper-to-lower and dark-to-lower polariton relaxation, whereas upper-to-dark polariton relaxation initially increases and then saturates beyond a certain number of layers. We find that our analytical description accurately captures our numerical results. Overall, our work identifies the key pathways of exciton-polariton relaxation and develops a general mechanistic and predictive theory for relaxation in realistic filled cavities or cavities coupled to material of finite thickness. 


\section { Data Availability}
    The data that support the plots within this paper and other findings of this study are available from the corresponding authors upon a reasonable request.
\section { Code Availability}

The source code that supports the findings of this study are available from the corresponding author upon reasonable request.
\section { Acknowledgments}
This work was supported by the Texas A\&M startup funds. This work used TAMU FASTER at the Texas A\&M University through allocations  PHY230021 and PHY240260 from the Advanced Cyberinfrastructure Coordination Ecosystem: Services \& Support (ACCESS) program, which is supported by National Science Foundation grants \#2138259, \#2138286, \#2138307, \#2137603, and \#2138296. L.B. is thankful for the support of the NSF-GRFP Grant DGE-2139772 and notes: ``This material is based upon work supported by the National Science Foundation Graduate Research Fellowship Program. Any opinions, findings, and conclusions or recommendations expressed in this material are those of the author(s) and do not necessarily reflect the views of the National Science Foundation.
\section{ Competing Interests}
The authors declare no competing interests.
\bibliography{bib.bib}

@article{OpalaOME2023,
  title={Harnessing exciton-polaritons for digital computing, neuromorphic computing, and optimization},
  author={Opala, Andrzej and Matuszewski, Micha{\l}},
  journal={Optical Materials Express},
  volume={13},
  number={9},
  pages={2674--2689},
  year={2023},
  publisher={Optica Publishing Group}
}

@article{SanvittoNM2016,
  title={The road towards polaritonic devices},
  author={Sanvitto, Daniele and K{\'e}na-Cohen, St{\'e}phane},
  journal={Nature Materials},
  volume={15},
  number={10},
  pages={1061--1073},
  year={2016},
  publisher={Nature Publishing Group UK London}
}

@article{ZasedatelevNP2019,
  title={A room-temperature organic polariton transistor},
  author={Zasedatelev, Anton V and Baranikov, Anton V and Urbonas, Darius and Scafirimuto, Fabio and Scherf, Ullrich and St{\"o}ferle, Thilo and Mahrt, Rainer F and Lagoudakis, Pavlos G},
  journal={Nature Photonics},
  volume={13},
  number={6},
  pages={378--383},
  year={2019},
  publisher={Nature Publishing Group UK London}
}

@article{AmoNP2010,
  title={Exciton--polariton spin switches},
  author={Amo, Alberto and Liew, TCH and Adrados, Claire and Houdr{\'e}, Romuald and Giacobino, Elisabeth and Kavokin, AV and Bramati, A},
  journal={Nature Photonics},
  volume={4},
  number={6},
  pages={361--366},
  year={2010},
  publisher={Nature Publishing Group UK London}
}

@article{MandalCR2023,
  title={Theoretical advances in polariton chemistry and molecular cavity quantum electrodynamics},
  author={Mandal, Arkajit and Taylor, Michael AD and Weight, Braden M and Koessler, Eric R and Li, Xinyang and Huo, Pengfei},
  journal={Chemical Reviews},
  volume={123},
  number={16},
  pages={9786--9879},
  year={2023},
  publisher={ACS Publications}
}

@article{basov2020polariton,
  title={Polariton panorama},
  author={Basov, Dmitri N and Asenjo-Garcia, Ana and Schuck, P James and Zhu, Xiaoyang and Rubio, Angel},
  journal={Nanophotonics},
  volume={10},
  number={1},
  pages={549--577},
  year={2020},
  publisher={De Gruyter}
}

@article{hutchison2012modifying,
  title={Modifying chemical landscapes by coupling to vacuum fields},
  author={Hutchison, James A and Schwartz, Tal and Genet, Cyriaque and Devaux, Elo{\"\i}se and Ebbesen, Thomas W},
  journal={Angewandte Chemie International Edition},
  volume={51},
  number={7},
  pages={1592--1596},
  year={2012},
  publisher={WILEY-VCH Verlag Weinheim}
}

@article{li2022molecular,
  title={Molecular polaritonics: Chemical dynamics under strong light--matter coupling},
  author={Li, Tao E and Cui, Bingyu and Subotnik, Joseph E and Nitzan, Abraham},
  journal={Annual review of physical chemistry},
  volume={73},
  number={1},
  pages={43--71},
  year={2022},
  publisher={Annual Reviews}
}

@article{GhoshAQT2021,
  title={Quantum neuromorphic computing with reservoir computing networks},
  author={Ghosh, Sanjib and Nakajima, Kohei and Krisnanda, Tanjung and Fujii, Keisuke and Liew, Timothy CH},
  journal={Advanced Quantum Technologies},
  volume={4},
  number={9},
  pages={2100053},
  year={2021},
  publisher={Wiley Online Library}
}

@article{BallariniNL2020,
  title={Polaritonic neuromorphic computing outperforms linear classifiers},
  author={Ballarini, Dario and Gianfrate, Antonio and Panico, Riccardo and Opala, Andrzej and Ghosh, Sanjib and Dominici, Lorenzo and Ardizzone, Vincenzo and De Giorgi, Milena and Lerario, Giovanni and Gigli, Giuseppe and others},
  journal={Nano Letters},
  volume={20},
  number={5},
  pages={3506--3512},
  year={2020},
  publisher={ACS Publications}
}

@article{MirekNL2021,
  title={Neuromorphic binarized polariton networks},
  author={Mirek, Rafa{\l} and Opala, Andrzej and Comaron, Paolo and Furman, Magdalena and Kr{\'o}l, Mateusz and Tyszka, Krzysztof and Seredy\'{n}ski, Bart{\l}omiej and Ballarini, Dario and Sanvitto, Daniele and Liew, Timothy CH and others},
  journal={Nano Letters},
  volume={21},
  number={9},
  pages={3715--3720},
  year={2021},
  publisher={ACS Publications}
}

@article{GhoshNPJQ2020,
  title={Quantum computing with exciton-polariton condensates},
  author={Ghosh, Sanjib and Liew, Timothy CH},
  journal={npj Quantum Information},
  volume={6},
  number={1},
  pages={16},
  year={2020},
  publisher={Nature Publishing Group UK London}
}

@article{BerloffNM2017,
  title={Realizing the classical XY Hamiltonian in polariton simulators},
  author={Berloff, Natalia G and Silva, Matteo and Kalinin, Kirill and Askitopoulos, Alexis and T{\"o}pfer, Julian D and Cilibrizzi, Pasquale and Langbein, Wolfgang and Lagoudakis, Pavlos G},
  journal={Nature Materials},
  volume={16},
  number={11},
  pages={1120--1126},
  year={2017},
  publisher={Nature Publishing Group UK London}
}

@article{RojasPRB2023,
  title={Topological Frenkel exciton polaritons in one-dimensional lattices of strongly coupled cavities},
  author={Rojas-S{\'a}nchez, J Andr{\'e}s and Jomaso, Yesenia A Garc{\'\i}a and Vargas, Brenda and Dom{\'\i}nguez, David Ley and Ordo{\~n}ez-Romero, C{\'e}sar L and Lara-Garc{\'\i}a, Hugo A and Camacho-Guardian, Arturo and Pirruccio, Giuseppe},
  journal={Physical Review B},
  volume={107},
  number={12},
  pages={125407},
  year={2023},
  publisher={APS}
}

@article{xiang2024molecular,
  title={Molecular polaritons for chemistry, photonics and quantum technologies},
  author={Xiang, Bo and Xiong, Wei},
  journal={Chemical Reviews},
  volume={124},
  number={5},
  pages={2512--2552},
  year={2024},
  publisher={ACS Publications}
}

@article{nagarajan2021chemistry,
  title={Chemistry under vibrational strong coupling},
  author={Nagarajan, Kalaivanan and Thomas, Anoop and Ebbesen, Thomas W},
  journal={Journal of the American Chemical Society},
  volume={143},
  number={41},
  pages={16877--16889},
  year={2021},
  publisher={ACS Publications}
}

@article{thomas2019tilting,
  title={Tilting a ground-state reactivity landscape by vibrational strong coupling},
  author={Thomas, Anoop and Lethuillier-Karl, Lucas and Nagarajan, Kalaivanan and Vergauwe, Robrecht MA and George, Jino and Chervy, Thibault and Shalabney, Atef and Devaux, Elo{\"\i}se and Genet, Cyriaque and Moran, Joseph and others},
  journal={Science},
  volume={363},
  number={6427},
  pages={615--619},
  year={2019},
  publisher={American Association for the Advancement of Science}
}

@article{ahn2023modification,
  title={Modification of ground-state chemical reactivity via light--matter coherence in infrared cavities},
  author={Ahn, Wonmi and Triana, Johan F and Recabal, Felipe and Herrera, Felipe and Simpkins, Blake S},
  journal={Science},
  volume={380},
  number={6650},
  pages={1165--1168},
  year={2023},
  publisher={American Association for the Advancement of Science}
}

@article{bhuyan2023rise,
  title={The rise and current status of polaritonic photochemistry and photophysics},
  author={Bhuyan, Rahul and Mony, Jurgen and Kotov, Oleg and Castellanos, Gabriel W and Gómez Rivas, Jaime and Shegai, Timur O and Borjesson, Karl},
  journal={Chemical Reviews},
  volume={123},
  number={18},
  pages={10877--10919},
  year={2023},
  publisher={ACS Publications}
}

@article{rashidi2025efficient,
  title={Efficient and tunable photochemical charge transfer via long-lived bloch surface wave polaritons},
  author={Rashidi, Kamyar and Michail, Evripidis and Salcido-Santacruz, Bernardo and Paudel, Yamuna and Menon, Vinod M and Sfeir, Matthew Y},
  journal={Nature Nanotechnology},
  pages={1--7},
  year={2025},
  publisher={Nature Publishing Group UK London}
}

@article{manjalingal2025tilted,
  title={Tilted Material in an Optical Cavity: Light-Matter Moir{\'e} Effect and Coherent Frequency Conversion},
  author={Manjalingal, Arshath and Rahmanian Koshkaki, Saeed and Blackham, Logan and Mandal, Arkajit},
  journal={ACS Photonics},
  year={2025},
  publisher={ACS Publications}
}

@misc{koshkaki2025,
      title={Exciton-Polariton Dynamics in Multilayered Materials}, 
      author={Saeed Rahmanian Koshkaki and Arshath Manjalingal and Logan Blackham and Arkajit Mandal},
      year={2025},
      eprint={2502.12933},
      archivePrefix={arXiv},
      primaryClass={quant-ph},
      url={https://arxiv.org/abs/2502.12933}, 
}

@article{Pandya_2022,
   title={Tuning the Coherent Propagation of Organic Exciton‐Polaritons through Dark State Delocalization},
   volume={9},
   ISSN={2198-3844},
   url={http://dx.doi.org/10.1002/advs.202105569},
   DOI={10.1002/advs.202105569},
   number={18},
   journal={Advanced Science},
   publisher={Wiley},
   author={Pandya, Raj and Ashoka, Arjun and Georgiou, Kyriacos and Sung, Jooyoung and Jayaprakash, Rahul and Renken, Scott and Gai, Lizhi and Shen, Zhen and Rao, Akshay and Musser, Andrew J.},
   year={2022},
   month=apr }

@article{hoffmann2020effect,
  title={Effect of many modes on self-polarization and photochemical suppression in cavities},
  author={Hoffmann, Norah M and Lacombe, Lionel and Rubio, Angel and Maitra, Neepa T},
  journal={The Journal of Chemical Physics},
  volume={153},
  number={10},
  year={2020},
  publisher={AIP Publishing}
}

@article{hoffmann2019capturing,
  title={Capturing vacuum fluctuations and photon correlations in cavity quantum electrodynamics with multitrajectory Ehrenfest dynamics},
  author={Hoffmann, Norah M and Sch{\"a}fer, Christian and Rubio, Angel and Kelly, Aaron and Appel, Heiko},
  journal={Physical Review A},
  volume={99},
  number={6},
  pages={063819},
  year={2019},
  publisher={APS}
}

@article{tichauer2021multi,
  title={Multi-scale dynamics simulations of molecular polaritons: The effect of multiple cavity modes on polariton relaxation},
  author={Tichauer, Ruth H and Feist, Johannes and Groenhof, Gerrit},
  journal={The Journal of Chemical Physics},
  volume={154},
  number={10},
  year={2021},
  publisher={AIP Publishing}
}

@article{yang2023enabling,
  title={Enabling multiple intercavity polariton coherences by adding quantum confinement to cavity molecular polaritons},
  author={Yang, Zimo and Bhakta, Harsh H and Xiong, Wei},
  journal={Proceedings of the National Academy of Sciences},
  volume={120},
  number={1},
  pages={e2206062120},
  year={2023},
  publisher={National Academy of Sciences}
}

@article{xu2023ultrafast,
  title={Ultrafast imaging of polariton propagation and interactions},
  author={Xu, Ding and Mandal, Arkajit and Baxter, James M and Cheng, Shan-Wen and Lee, Inki and Su, Haowen and Liu, Song and Reichman, David R and Delor, Milan},
  journal={Nature communications},
  volume={14},
  number={1},
  pages={3881},
  year={2023},
  publisher={Nature Publishing Group UK London}
}

@article{chng2025quantum,
  title={Quantum dynamics simulations of exciton polariton transport},
  author={Chng, Benjamin XK and Mondal, M Elious and Ying, Wenxiang and Huo, Pengfei},
  journal={Nano Letters},
  volume={25},
  number={4},
  pages={1617--1622},
  year={2025},
  publisher={ACS Publications}
}

@article{ying2025microscopic,
  title={Microscopic theory of polariton group velocity renormalization},
  author={Ying, Wenxiang and Chng, Benjamin XK and Delor, Milan and Huo, Pengfei},
  journal={Nature Communications},
  volume={16},
  number={1},
  pages={6950},
  year={2025},
  publisher={Nature Publishing Group UK London}
}

@article{krupp2025quantum,
  title={Quantum dynamics simulation of exciton-polariton transport},
  author={Krupp, Niclas and Groenhof, Gerrit and Vendrell, Oriol},
  journal={Nature Communications},
  volume={16},
  number={1},
  pages={5431},
  year={2025},
  publisher={Nature Publishing Group UK London}
}

@article{sokolovskii2025disentangling,
  title={Disentangling enhanced diffusion and ballistic motion of excitons coupled to Bloch surface waves with molecular dynamics simulations},
  author={Sokolovskii, Ilia and Luo, Yunyi and Groenhof, Gerrit},
  journal={The Journal of Physical Chemistry Letters},
  volume={16},
  number={26},
  pages={6719--6727},
  year={2025},
  publisher={ACS Publications}
}

@article{blackham2025microscopic,
  title={Microscopic theory of polaron-polariton dispersion and propagation},
  author={Blackham, Logan and Manjalingal, Arshath and Koshkaki, Saeed Rahmanian and Mandal, Arkajit},
  journal={Nano Letters},
  year={2025},
  publisher={ACS Publications}
}

@article{ghosh2025mean,
  title={Mean-field mixed quantum-classical approach for many-body quantum dynamics of exciton polaritons},
  author={Ghosh, Pritha and Manjalingal, Arshath and Wickramasinghe, Sachith and Koshkaki, Saeed Rahmanian and Mandal, Arkajit},
  journal={Physical Review B},
  volume={112},
  number={10},
  pages={104319},
  year={2025},
  publisher={APS}
}

@article{perez2025radiative,
  title={Radiative pumping vs vibrational relaxation of molecular polaritons: a bosonic mapping approach},
  author={P{\'e}rez-S{\'a}nchez, Juan B and Yuen-Zhou, Joel},
  journal={Nature Communications},
  volume={16},
  number={1},
  pages={3151},
  year={2025},
  publisher={Nature Publishing Group UK London}
}

@article{wang2025robust,
  title={Robust Surface-Induced Enhancement of Exciton Transport in Magic-Angle-Oriented Molecular Aggregates},
  author={Wang, Siwei and Hsu, Liang-Yan and Chen, Hsing-Ta},
  journal={The Journal of Physical Chemistry Letters},
  volume={16},
  pages={10575--10583},
  year={2025},
  publisher={ACS Publications}
}

@article{catuto2025interplay,
  title={Interplay between static and dynamic disorder: Contrasting effects on dark state population inside a cavity},
  author={Catuto, Robert F and Chen, Hsing-Ta},
  journal={The Journal of Chemical Physics},
  volume={162},
  number={20},
  year={2025},
  publisher={AIP Publishing}
}

@article{AroeiraJCP2025,
    author = {Aroeira, Gustavo J. R. and Ribeiro, Raphael F.},
    title = {Static disorder-induced renormalization of polariton group velocity},
    journal = {The Journal of Chemical Physics},
    volume = {163},
    number = {12},
    pages = {124120},
    year = {2025},
    month = {09}}

@article{mandal2023microscopic,
  title={Microscopic theory of multimode polariton dispersion in multilayered materials},
  author={Mandal, Arkajit and Xu, Ding and Mahajan, Ankit and Lee, Joonho and Delor, Milan and Reichman, David R},
  journal={Nano Letters},
  volume={23},
  number={9},
  pages={4082--4089},
  year={2023},
  publisher={ACS Publications}
}

@Article{Yongseok2025,
author={Hong, Yongseok
and Xu, Ding
and Delor, Milan},
title={Exciton delocalization suppresses polariton scattering},
journal={Chem},
year={2025}
}

@article{polak2020manipulating,
  title={Manipulating molecules with strong coupling: harvesting triplet excitons in organic exciton microcavities},
  author={Polak, Daniel and Jayaprakash, Rahul and Lyons, Thomas P and Mart{\'\i}nez-Mart{\'\i}nez, Luis {\'A} and Leventis, Anastasia and Fallon, Kealan J and Coulthard, Harriet and Bossanyi, David G and Georgiou, Kyriacos and Petty, Anthony J and others},
  journal={Chemical science},
  volume={11},
  number={2},
  pages={343--354},
  year={2020},
  publisher={Royal Society of Chemistry}
}

@article{pandya2022tuning,
  title={Tuning the coherent propagation of organic exciton-polaritons through dark state delocalization},
  author={Pandya, Raj and Ashoka, Arjun and Georgiou, Kyriacos and Sung, Jooyoung and Jayaprakash, Rahul and Renken, Scott and Gai, Lizhi and Shen, Zhen and Rao, Akshay and Musser, Andrew J},
  journal={Advanced Science},
  volume={9},
  number={18},
  pages={2105569},
  year={2022},
  publisher={Wiley Online Library}
}

@article{Lydick:24,
author = {Nathanial Lydick and Jiaqi Hu and Hui Deng},
journal = {J. Opt. Soc. Am. B},
number = {8},
pages = {C247--C253},
publisher = {Optica Publishing Group},
title = {Dimensional dependence of a molecular-polariton mode number},
volume = {41},
month = {Aug},
year = {2024}

}

@article{sun2025exploring,
  title={Exploring the Delocalization of Dark States in a Multimode Optical Cavity},
  author={Sun, Kunyang and Du, Matthew and Yuen-Zhou, Joel},
  journal={The Journal of Physical Chemistry C},
  volume={129},
  number={21},
  pages={9837--9843},
  year={2025},
  publisher={ACS Publications}
}

@article{keeling2020bose,
  title={Bose--Einstein condensation of exciton-polaritons in organic microcavities},
  author={Keeling, Jonathan and K{\'e}na-Cohen, St{\'e}phane},
  journal={Annual Review of Physical Chemistry},
  volume={71},
  number={1},
  pages={435--459},
  year={2020},
  publisher={Annual Reviews}
}

@article{neuman2018origin,
  title={Origin of the asymmetric light emission from molecular exciton--polaritons},
  author={Neuman, Tom{\'a}{\v{s}} and Aizpurua, Javier},
  journal={Optica},
  volume={5},
  number={10},
  pages={1247--1255},
  year={2018},
  publisher={Optical Society of America}
}

@article{groenhof2019tracking,
  title={Tracking polariton relaxation with multiscale molecular dynamics simulations},
  author={Groenhof, Gerrit and Climent, Claudia and Feist, Johannes and Morozov, Dmitry and Toppari, J Jussi},
  journal={The journal of physical chemistry letters},
  volume={10},
  number={18},
  pages={5476--5483},
  year={2019},
  publisher={ACS Publications}
}

@article{ChowdhuryJCP2021,
  title={Ring polymer quantization of the photon field in polariton chemistry},
  author={Chowdhury, Sutirtha N and Mandal, Arkajit and Huo, Pengfei},
  journal={The Journal of Chemical Physics},
  volume={154},
  number={4},
  year={2021},
  publisher={AIP Publishing}
}

@article{troisi2003rate,
  title={A rate constant expression for charge transfer through fluctuating bridges},
  author={Troisi, Alessandro and Nitzan, Abraham and Ratner, Mark A},
  journal={The Journal of chemical physics},
  volume={119},
  number={12},
  pages={5782--5788},
  year={2003},
  publisher={American Institute of Physics}
}

@article{beratan2009steering,
  title={Steering electrons on moving pathways},
  author={Beratan, David N and Skourtis, Spiros S and Balabin, Ilya A and Balaeff, Alexander and Keinan, Shahar and Venkatramani, Ravindra and Xiao, Dequan},
  journal={Accounts of chemical research},
  volume={42},
  number={10},
  pages={1669--1678},
  year={2009},
  publisher={ACS Publications}
}
\bibliographystyle{naturemag}

\end{document}